\documentclass[12pt,preprint]{aastex}
\usepackage{graphicx,verbatim}

\slugcomment{Draft}

\shorttitle{Charge States in CMEs}
\shortauthors{Rakowski, Laming, \& Lepri}

\begin{document}

\title{Ion Charge States in Halo CMEs: What can we Learn about the Explosion?}


\author{Cara E. Rakowski\altaffilmark{1,2}, J. Martin Laming\altaffilmark{2}
\& Susan T. Lepri\altaffilmark{3}}


\altaffiltext{1}{NRL/NRC Research Associate}
\altaffiltext{2}{E. O Hulburt Center for Space Research, Naval
Research Laboratory, Code 7674L, Washington DC 20375-5321
\email{crakowski@ssd5.nrl.navy.mil, laming@nrl.navy.mil}}
\altaffiltext{3}{Department of
Atmospheric, Oceanic and Space Sciences, University of Michigan, Ann
Arbor, MI 48109-2143\email{slepri@umich.edu}}

\begin{abstract}
We describe a new modeling approach to develop a more quantitative
understanding of the charge state distributions of the ions of
various elements detected in situ during halo Coronal Mass Ejection
(CME) events by the Advanced Composition Explorer (ACE) satellite.
Using a model CME hydrodynamic evolution based on observations of
CMEs propagating in the plane of the sky and on theoretical models,
we integrate time dependent equations for the ionization balance of
various elements to compare with ACE data. We find that plasma in
the CME ``core'' typically requires further heating following
filament eruption, with thermal energy input similar to the kinetic
energy input. This extra heating is presumably the result of post
eruptive reconnection. Plasma corresponding to the CME ``cavity'' is
usually not further ionized, since whether heated or not, the low
density gives freeze-in close the the Sun. The current analysis is
limited by ambiguities in the underlying model CME evolution. Such
methods are likely to reach their full potential when applied to
data to be acquired by STEREO when at optimum separation. CME
evolution observed with one spacecraft may be used to interpret CME
charge states detected by the other.
\end{abstract}



\section{Introduction}
Coronal Mass Ejections (CMEs) represent perhaps the most extreme
manifestation of solar activity. Of order $10^{16}$ g of plasma is
expelled at speeds of several hundred m s$^{-1}$ to in excess of one
thousand km s$^{-1}$. Such speeds are frequently super-Alfv\'enic
with respect to the ambient solar wind, and the shocks driven by CME
events can be efficient accelerators of energetic particles,
constituting the main radiation hazard for space-borne
instrumentation. The spectacular nature of the
phenomenon, and its relevance to space based technology, have
spawned much theoretical and observational work with the ultimate
goal of understanding CMEs sufficiently deeply to enable the
forecasting of such events, in a discipline that has become known as
``space weather''.

The interplanetary manifestation of CMEs (ICMEs) embedded in the
solar wind can exhibit a variety of signatures in the magnetic
field, solar wind speed and density profiles, proton thermal
properties, elemental and ionic composition, and the presence of
energetic particles (see Zurbuchen and Richardson, 2006 for more
detail). Of particular interest here are the ionic charge states
observed in the solar wind during ICMEs.  Unlike many plasma
properties, the charge states are determined within 4 solar radii
and remain unchanged as they expand further into the heliosphere.
This property makes them an important signature of the eruption
process close to the Sun. A large body of literature exists
highlighting the rich data sets which detail the unique charge
composition observed within ICMEs (e.g. Zurbuchen and Richardson
2006, Zurbuchen et al. 2004, Lepri and Zurbuchen 2004, Lepri et al.
2001, Henke et al. 2001, Gloeckler et al. 1998, Henke et al. 1998,
Galvin \& Gloeckler 1997, Reinard 2005).

While a number of studies have examined the existing solar wind
composition data and made inferences on freeze-in temperatures based
on computed ionization distributions appropriate to coronal
equilibrium, in this paper we commence time-dependent modeling of
the ion charge state distributions of various elements to
draw quantitative conclusions regarding the thermal energy input and
initial conditions in the corona during the CME eruption. Using
current ionization-recombination calculations and a simple model for
the spatial and temperature evolution of the CME plasma we attempt
to reproduce the charge states detected {\it in situ} by instruments
on the Advanced Composition Explorer (ACE) spacecraft. In order to
provide some context for our work, we first review the status of
theory and observation for CMEs. Following this, we describe our
modeling approach in some detail, and apply our methods to a sample
of ICMEs detected ACE, which were chosen to provide a variety of ICME
speeds and charge state distributions.

\section{CME Theory and Observation: Current Status}

CMEs were first discovered during the Skylab era \citep{gosling74}, and a ``halo CME'' (i.e. an Earth
directed event) was first detected by the P78-1 coronagraph \citep{howard85}. Since the 1995 launch of
SOHO, they are now routinely observed and studied by the Extreme Ultraviolet Imaging Telescope (EIT)
and Large Angle Spectroscopic Coronagraph (LASCO) instruments \citep[e.g.][]{dere97}.
LASCO white light observations are sensitive to photospheric radiation Thomson scattered by
free electrons in the
corona. Three nested coronagraphs provided coverage between 1.5 and 30 $R_{\sun}$ heliocentric distance,
reduced to  2.5 - 30 $R_{\sun}$ following the 1998 demise of C1.
EIT records images in narrow EUV wavebands emitted closer to the solar surface, and allows investigations
of the coronal precursor and response to the CME eruption. This instrumentation provides the context
for the observations of most importance in this work, those of particle charge states and masses detected
{\it in situ} at the L1 Lagrange point by mass and charge to mass spectrometers on board the
Advanced Composition Explorer (ACE). We use Level 2 data supplied by the ACE Science Center from
the SWICS/SWIMS (Solar Wind Ion Composition Spectrometer/Solar Wind Ions Mass Spectrometer) and
SWEPAM (Solar Wind Electron Proton Alpha Monitor) instruments.

For our purposes, the most important feature of CMEs is their rate
of expansion and acceleration close to the Sun. For the time being,
we are not able to simultaneously observe the expansion rate of a
CME for which we also detect the particle emission, and must
estimate the velocity profile of a halo CME from observations of
other CMEs propagating in the plane of the sky. Happily, these
frequently follow a similar form \citep{zhang01,zhang04,zhang06}.
Observationally, CME evolution can be divided into three phases. An
initial phase of expansion at approximately constant velocity in the
range 10-100 km s$^{-1}$, is called the ``initiation phase'', during
which the flux rope rises to a height of about 1.5 $R_{\sun}$.
This  is followed by the ``acceleration phase''. During
this period, the CME undergoes a roughly constant acceleration up to
its final speed, between a few 100 km s$^{-1}$ to a few 1000 km
s$^{-1}$ for the fastest CMEs. Typically this final speed is
achieved around 5-10 $R_{\sun}$. Lastly, the
``propagation phase'' with essentially constant velocity transports
the CME to 1 AU and beyond. These features are also apparent in
theoretical work. In the breakout model
\citep{antiochos98,antiochos99,devore00,aulanier02}, the initiation
phase corresponds to the initial shearing of the magnetic field
lines closest to the filament channel in an overlying arcade of
loops. As the filament rises and reconnection commences underneath
it, the acceleration phase begins, typically at 1.5 $R_{\sun}$
\citep[see Fig. 1 in][]{lynch04}. In the breakout model,
posteruptive reconnection beneath the erupting filament is not
strictly necessary for an explosion to occur, though reconnection
above the filament is required. There appears to be no clear event
to signal the end of the acceleration phase, though clearly a CME
cannot continue accelerating forever. Analytic models of CME
eruptions \citep{lin00} can also give similar height-time and
velocity-time profiles. Again, the transition from the initiation
phase to the acceleration phase corresponds to the formation of a
current sheet below the flux rope. The resulting analytic solutions
for the height and velocity of the reconnection-driven flux rope do
indeed show an approximately constant acceleration phase, followed
by a roughly constant velocity phase, at least for models with
relatively high reconnection rates producing fast CMEs. Some
deceleration may also occur during the propagation phase.

The morphology of the erupting CME plasma, while generally quite
variable, shows a regular pattern of features for flux rope or
magnetic cloud CMEs, which are thought to be associated with
erupting prominences, and which will be the focus of most of our
modeling using the event list of \citet{lynch06}. The CME front (a
shock front in CMEs fast enough to drive a shock) encloses a
``cavity'' region, presumably a region of strong magnetic field
comprising the flux rope \citep{lynch04}. Towards the rear of the
erupting plasma is the CME ``core'', and region with density perhaps
a factor of 10 higher than in the cavity. This region possibly has
an origin as the cold prominence material. Reconnection above the
erupting filament in the breakout model would be expected to heat
the cavity plasma, if anything. Unfortunately the cavity plasma is
usually too rarefied for any heating to be visible in the detected
ion charge states. Reconnection behind the eruption
\citep[e.g.][]{riley02,riley07}, for example to form postflare
loops, will most likely heat the core material, where the density is
high enough that increased ionization can set in before freeze-in.
\citet{kumar96} model a lower velocity flux rope eruption, making
the assumption that magnetic helicity is conserved during the
process. They find that the magnetic energy of the flux rope must
decrease, and that not all of this energy can be converted into
kinetic energy of the expanding plasma, some must go to heat, though
the exact form and location of the magnetic energy dissipation is
not specified.

The flux rope may also be distorted during transit to 1 AU. \citet{riley04} model the effects of
spherical expansion and the effect of pressure gradients between the CME plasma and the ambient solar
wind. The flux rope may expand substantially in latitude. \citet{schmidt01} consider the case of the
flux rope propagating in the velocity shear layer between the fast and slow solar wind, and find that
the magnetic topology may be changed by reconnection. In this way, plasma from the CME ``core'' and
``cavity'' regions may become mixed together before {\it in situ} detection at 1 AU by ACE.

\section{Modeling Approach}

Our simulations follow the evolution of high density core and lower density cavity material for a CME.
The ionization balance for the two components are followed separately starting from  coronal
or flare temperatures allowing for heating during the acceleration phase of the CME evolution,
particularly of the core material. Various parameters listed below are adjusted to match the velocity,
plasma density, and most importantly the charge state distributions of
O, Si and Fe observed by ACE for a small selection of CMEs.
Given the degeneracies in parameter
space that will be removed by the simultaneous observation of CMEs from face-on and edge-on viewpoints
with STEREO, our primary intent at this time is to illustrate that the CME charge states can be successfully
modeled under reasonable assumptions.

\subsection{Charge State Evolution}

Our approach to modeling CME charge states is to follow the behavior of the ionization
balance of a Lagrangian plasma packet, using an analytic prescription for the hydrodynamic or
magnetohydrodynamic evolution. We use an adaptation of the BLASPHEMER (BLASt Propagation
in Highly EMitting EnviRonment)\footnote{The name gives away its origin in modeling
laboratory and astrophysical shock waves.} code
\citep{laming02,laming03g,laming03,laming04},
which follows the time dependent
ionization balance and temperatures of a Lagrangian plasma parcel as it
expands in the solar wind. The initial conditions are set by assuming ionization
equilibrium at an electron temperature of $1-3\times 10^{6}$K appropriate for coronal plasma.

The density $n_{iq}$ of ions of element $i$ with charge $q$ is
given by
\begin{equation} {dn_{iq}\over dt} =
n_e\left(C_{ion,q-1}n_{i~q-1}-C_{ion,q}n_{iq}\right)+
n_e\left(C_{rr,q+1} +C_{dr,q+1}\right)n_{i~q+1}-
n_e\left(C_{rr,q}+ C_{dr,q}\right)n_{iq}
\end{equation}
where $C_{ion,q}, C_{rr,q}, C_{dr,q}$ are
the rates for electron impact ionization, radiative recombination and dielectronic
recombination respectively, out of the charge
state $q$. These rates are the same as those used in the recent ionization balance
calculations of \citet{mazzotta98}, using subroutines kindly supplied by
Dr P. Mazzotta (private communication 2000), with the following updates.
Dielectronic recombination from H- to He-like and from He- to Li-like are
taken from \citet{dasgupta04}. Dielectronic recombination for the successive isoelectronic sequences Li-,
Be-, B-, C-, N-, O-, and F-like are taken from \citet{colgan04},
\citet{colgan03}, \citet{altun04}, \citet{zatsarinny04a}, \citet{mitnik04},
\citet{zatsarinny03}, and \citet{gu03} respectively. Additionally, dielectronic
recombination from Ne- to Na-like and from Na- to Mg-like are taken from
\citet{zatsarinny04b} and \citet{gu04}. We also take dielectronic recombination rates for Fe$^{13+}$
from \citet{badnell06}. The electron density $n_e$ is determined
from the condition that the plasma be electrically neutral. The ion and electron
temperatures, $T_{iq}$ and
$T_e$ are coupled by Coulomb collisions by
\begin{equation} {dT_{iq}\over dt}= -0.13n_e{\left(T_{iq}-T_e\right)\over M_{iq}T_e^{3/2}}
{q^3n_{iq}/\left(q+1\right)\over\left(\sum _{iq} n_{iq}\right)}\left(\ln\Lambda\over 37\right)
\end{equation} and
\begin{equation}
{dT_e\over dt}= {0.13n_e\over T_e^{3/2}}\sum _{iq}{\left(T_{iq}-T_e\right)\over M_{iq}}
{q^2n_{iq}/\left(q+1\right)\over\left(\sum _{iq} n_{iq}\right)}\left(\ln\Lambda\over 37\right)
-{T_e\over n_e}\left({dn_e\over dt}\right)_{ion} - {2\over 3n_ek_{\rm B}}
{dQ\over dt}.
\end{equation}
Here $M_{iq}$ is the atomic mass of the ions of element $i$ and
charge $q$ in the plasma,
and $\ln\Lambda\simeq 28$ is the Coulomb logarithm. The term in $dQ/dT$
represents plasma energy losses due to ionization and radiation.
The term $-\left(T_e/n_e\right)\left(dn_e/dt\right)_{ion}$ gives the reduction
in electron temperature when the electron density increases due to ionization.
Recombinations, which reduce the electron density, do not result in an increase
in the electron temperature in low density plasmas, since the energy of the
recombined electron is radiated away (in either radiative or dielectronic recombination).

\subsection{Hydrodynamic Evolution}

We based our model CME evolution on the phenomenology described in section 2, of an initiation, acceleration
and propagation phase, to use the terminology of \citep{zhang01,zhang04}.
Plasma starts at $1.5 R_{\sun}$ moving at an initial velocity, $v_{i}$,
between 10 and 100 km s$^{-1}$, i.e. as it moves from the ``initiation'' to the ``acceleration'' phase.
The initial electron density, is taken to be either near $10^8$ cm$^{-3}$ or a factor of 10 lower,
corresponding to plasma in the
CME ``core'' or ``cavity'' respectively.
An acceleration, $a$, of 0.1---0.5 km s$^{-2}$ is chosen so that the
CME reaches its final coasting velocity, $v_{f}$, at a heliocentric distance of 3---10$R_{\sun}$ ($R_{c}$). The
density is assumed to fall off as $1/r^{\left(2+v_A/\left(v_A+v_r\right)\right)}$, where $v_A$ is the
coronal Alfv\'en speed and $v_r$ is the CME expansion speed.
This form, with a suitable choice for $v_A$ ($\sim 1000$ km s$^{-1}$), gives an initial superradial
expansion as the CME expands laterally, going over to a $1/r^2$ form at large distances from the Sun.
The Alfv\'en speed, $v_A$, is assumed to vary approximately as $1/r^{1/3}$ or $n_e^{1/6}$, coming from
the approximate relations $B\propto 1/r^{4/3}$ (the geometric mean of $B_r\propto 1/r^2$ and
$B_{\theta}\propto B_{\phi}\propto 1/r$) and $\rho\propto 1/r^2$. \citet{mann03} give a more detailed
account of the variation of the Alfv\'en speed with radius in the quiescent solar corona above an active region.
However our simpler form gives a density dependence on radius which matches
very well with the various plots in \citet{lynch04}, and is certainly adequate for our needs.


For simplicity we assume the core and cavity differ in density by a
factor of 10, but contribute roughly equal amounts of material by
mass and undergo the same velocity and expansion evolution. Their
electron temperatures at ``initiation'' are either assumed to be the
same or the cavity temperature is held lower, in the range of
typical coronal, rather than flare, temperatures. Subsequent heating
of the higher density core plasma during the acceleration phase is
explored assuming a heating rate for the CME plasma proportional to
the rate of kinetic energy increase, i.e. a constant fraction
$QE/KE$ during the acceleration up to $v_{f}$. Here we assume that
all particle species are heated equally, but our models are only
sensitive to the electron temperature. Hence depending on how the
magnetic reconnection partitions energy between electrons and ions,
and whether the heat input is constant during the acceleration
phase, different CME energy budgets may result. Having the heating
be proportional to the acceleration during the acceleration phase is
one choice which allows for particularly easy comparisons of the
energetics. However, we cannot distinguish between this scenario and
impulsive heating to high temperatures (2---3$\times 10^{7}$K)
during the initiation phase (with insufficient time to reach
ionization equilibrium). The observation of high charge states
simply requires that high temperatures be reached while the density
is still high enough to allow for significant ionization of the high
Z elements. Such heating was not further explored for the cavity
material simply because at those densities it would have little or
no impact on the ionization state which freezes in well below
2$R_{\sun}$.

\section{Modeling a Sample of ACE Coronal Mass Ejections}

A sample of 6 ICME events was selected from a catalog of magnetic
cloud events by \citet{lynch06}, supplemented by two more events
from the survey of \citet{ugarte07}, all listed in Table 1. Listed
here are the average, standard deviation and maxima of the He
velocity, proton density and the abundance ratios of He/O and Fe/O
as measured at ACE with the SWICS/SWIMS and SWEPAM instruments
(proton density). The abundance ratios seen were generally typical
for coronal material except for the Halloween 2003 event which shows
significantly enhanced Fe/O. He/O ratios on the other hand,
especially in the faster CMEs, are more typical of the chromospheric
value of $\sim 130$, or of that found in flares \citep{feldman05},
and not the lower values found elsewhere in the solar wind and
corona \citep[e.g.][]{lamingf03,kasper07}. The observed velocity may
be considered a lower limit, since deceleration is likely to have
occurred in transit to 1 AU, however our simulation results were
relatively insensitive to the final velocity. The proton density at
1 AU can be compared to the electron density at 10$R_{\sun}$ by
multiplying by a factor of $\sim400$ for the $1/r^{2}$ density fall
off and another factor of 1.2---1.6 for the proton to electron
density ratio. We expect the measured proton density to reflect some
combination of core and cavity material.

Examples of successful CME models for the chosen events are listed in Table 2 in order of increasing velocity.
The dominant charge states for each model are listed in order of abundance composing a total
charge fraction of at least 0.8.
These are representative of the charge states observed at ACE, which, however, did vary over the
course of an ICME event.   Examples of the observed charge states at different slices in time,
the final modeled charge states of the combined core and cavity plasma, and the initial evolution
of ion fractions in the core material are shown for the 2001 doy 351 ICME and the Halloween 2003
events in figures 1, 2,and 3 and 4, 5, and 6, respectively.
The model solution for any given event is not unique, and we used this freedom to choose an ensemble that
exhibits the range of behavior possible and the effect of any given parameter.
To begin modeling, the parameters of the initial velocity, density, acceleration and lateral expansion
(via the Alfv\'en speed) are chosen to match the velocity and plasma density of the CME in question as
measured by ACE. The model acceleration and expansion also agree well with observational
results in \citet{vourlidas03} and \citet{thernisien06} for the region behind the forward shock.
Both theoretical models and observations of CMEs in the plane of the sky show a high density core of
plasma surrounded by a lower density cavity.  The bimodal charge distribution of both Fe and Si in most
of the CMEs considered here is naturally concordant with a mixing of cavity and core material
during the CME flux rope passage to 1 AU after freeze-in has occurred.

In 5 out of 8 sample CMEs the dominant Fe charge states are Neon-like (16+) or higher,
indicating that high temperatures,
comparable to flares ($\sim 10^7$K), are involved. Starting the plasma from this temperature and allowing the
ions to recombine as they expand can often account for the Fe ionization balance, with peaks around Fe$^{16+}$
and Fe$^{8+}$ (the Ne-like and Ar-like charge states, which have small recombination rates to the next charge
states down, hence population ``bottlenecks'' here). However the lower-Z elements place a limit on the maximum
starting temperature (at least if assuming ionization equilibrium in the seed plasma).
Above $\sim 2.5\times 10^{6}$K
O would be mainly O$^{8+}$ instead of O$^{7+}$ and O$^{6+}$ as observed and would not recombine significantly
during the CME evolution.
Evidently plasma must start out much cooler, and be further heated as the CME accelerates. While various
possibilities exist among CME models regarding the role of reconnection in initiating or accelerating the
CME, nearly all of them require reconnection to produce the post-eruptive arcades of magnetic loops
\citep[e.g.][]{riley07}.
We follow \citet{lepri04} and argue that this reconnection must also heat the CME plasma
and that this is the cause for the high charge states.
Of the 3 events without a peak at high charge states in Fe, in ICMEs 2003 doy 129 and 2001 doy 351
the broad distribution of
both Si and Fe does require some heating. In the slowest event that shows no elevated charge states,
2002 doy 173, no heating is required but neither can it be excluded.

Finally, in Table 2 we also enter for some CMEs the velocity of the
CME front as deduced from SOHO/LASCO height-time
observations\footnote{using the CME catalog at
http://cdaw.gsfc.nasa.gov/CME\_list }, and the acceleration
necessary to produce this velocity at the heights observed. The
variation in heating rate as a multiple of the CME kinetic energy is
surprisingly small. Only for the extreme 2003 October 29 (doy 302)
event is the modeling really sensitive to this behavior. We caution
that the height-time observations give the velocity of the CME
front, not the expansion velocity of the flux rope or magnetic cloud
containing the ejecta, and that it is not possible in all cases to
unambiguously identify the disk event that gave rise to the observed
ICME.

\section{Discussion and Conclusions}
Our basic result, that the erupting CME plasma must be further heated as it expands, is not entirely
unexpected from theoretical considerations. \citet{kumar96} model the evolution of an expanding flux rope.
For a ratio of initial gravitational energy to magnetic energy in the flux rope of 0.22, they find
that between 0.58 and 0.78 of the initial magnetic energy is converted to forms of energy other the
gravitational potential energy and kinetic energy of the plasma, i.e., is available for plasma heating
or the generation of radiation or plasma waves. The maximum amount of magnetic energy that can go to
plasma kinetic energy is 0.38, so {\it at least} 1.5 times the kinetic energy in their model must appear as
heat.

\citet{kumar96} do not specify the precise mechanism by which the
magnetic energy is converted into other forms, arguing as they do
simply from conservation of total energy and magnetic helicity. On
the other hand, \citet{lin00} considered in a little more detail the
expulsion of a current carrying flux rope, paying more attention to
the initial quasi-equilibrium configuration and specifying magnetic
reconnection as the mechanism of releasing magnetic energy. Their
treatment of reconnection assumes, somewhat arbitrarily, that all
magnetic energy destroyed reappears as plasma kinetic energy, and
hence the CME speed in their model is therefore an overestimate.
However they do comment that previous numerical simulations by
\citet{forbes91} show that ``only about half of the released
magnetic energy is actually converted into the kinetic energy of the flux
rope; the other half goes into heating, radiation, and the
generation of plasma waves''. Therefore these authors also suggest
that an amount of thermal energy, similar to the kinetic energy of
the CME, should  be generated, and that this should result from
magnetic reconnection. However magnetic reconnection does not power
the CME. It is accelerated by ${\bf j} \times {\bf B}$ forces within
the flux rope, and reconnection eliminates magnetic field that holds
the flux rope down in the initial quasi-equilibrium configuration.

Magnetic reconnection plays a similar role in the breakout model of CMEs \citep{antiochos98,antiochos99,
devore00,aulanier02}.  Again, it plays the role of destroying magnetic field that otherwise holds
the CME plasma in equilibrium. Once this magnetic field is eliminated, the CME is powered by the
magnetic field below, stressed by shear motions either side of the neutral line. Reconnection at
this lower site, which forms postflare loops, has nothing to do with allowing the CME to erupt, but
may still be a heat source for the CME plasma. Hence in both the \citet{lin00} flux rope model and
in the breakout scenario, no strong prediction exists for how much plasma heating should be
expected as a function of CME kinetic energy, though simple estimates of the magnetic energy destroyed
in these cases are indeed of the right order of magnitude to supply the amount of heat we see
in the charge state distributions. \citet{shiota05} give a numerical simulation of a flux rope ejection,
and stress the role of the slow mode MHD shock in providing heat to the flux rope plasma.

Our estimates of post-eruptive heating are of course sensitive to the electron heating only, although we
assume that all plasma particles are similarly heated. Observations of post eruption electron acceleration
are reported by \citet{klassen05} for the 2003 October 28 event. Type III, II, and IV radio bursts are detected
coincident with the soft X-ray rise of the flare. Following that, an impulsive electron event in the energy
range 0.027-0.182 MeV and a gradual electron event with energies 0.31-10.4 MeV were detected by WIND-3DP
and SOHO/COSTEP, with a total duration of about 30 minutes following CME onset, by which time the CME
front had expanded to about 6$R_{\sun}$. The total energy in these accelerated electrons is significantly
less than the heating requirements we give in Table 2 \citep{mewaldt05}, however the timescale following
eruption is commensurate with our requirements. The expansion velocity of the CME front given by
\citet{klassen05}, derived from height-time measurements from SOHO/LASCO observations, is also
higher than the CME velocity we give in Table 2, and requires a significantly larger acceleration to
achieve this velocity by about 5.8$R_{\sun}$ where it was first observed by SOHO/LASCO. We have
experimented with different accelerations, and find that for this particular CME, for accelerations
above 1-2 km s$^{-2}$, it is impossible to match the observed charge states for any post eruption heating
rate. We emphasize that we are principally interested in the plasma in the CME ``core'', and that the
expansion of this material might not be well represented by the SOHO/LASCO observations of the CME front.
Additionally, this was a halo CME, so SOHO/LASCO observes the CME front off to one side of the main
eruption. We anticipate that such ambiguities will be substantially resolved with the advent of STEREO data.

Similar conclusions to ours about the thermal energy input to CMEs
have been reached from analysis of ultra-violet spectra taken by
SOHO/UVCS. \citet{akmal01} studied a 480 km s$^{-1}$ CME observed on
1999 April 23, and found a thermal energy comparable to the bulk
kinetic energy of the plasma. \citet{ciaravella01} gave similar
results for the 260 km s$^{-1}$ 1997 December 12 CME, while
\citet{lee07} studied the 2001 December 13 event, also examined by
\citet{ugarte07}, and included in our Table 2. None of these appear
in the magnetic cloud event list of \citet{lynch06}.

To summarize, our work on interpreting charge state distributions for the ions of various elements support
previous ideas in the literature that CME plasma continues to be heated as the eruption proceeds. A
future quantitative study of this may yield novel insights into the mechanism(s) of explosion, especially
because the charge states are formed by processes close to the Sun, and are then transported unchanged
to 1 AU.

\acknowledgements
This work has been supported by NASA LWS Grant NNH05AA05I (CER \& JML) and by an NSF SHINE Postdoctoral
Fellowship (STL). It made use of the CME catalog generated and maintained at the CDAW Data Center
by NASA and The Catholic University of America in cooperation with the Naval Research Laboratory.
SOHO is a project of international cooperation between ESA and NASA. This work has also made use of
ACE Level 2 SWICS and SWEPAM data provided by the ACE Science Center.


\clearpage
\begin{table}[t]
\begin{center}
\caption{ICME Observations}
\begin{tabular}{@{}*{10}{c}}  
\tableline\tableline \\
Year & Start
& \multicolumn{2}{c}{$v_{\rm He++}$ (km s$^{-1}$)}\tablenotemark{a}
& \multicolumn{2}{c}{$\rho_{\rm H+}$ (cm$^{-3}$)}
& \multicolumn{2}{c}{He/O\tablenotemark{b}}
& \multicolumn{2}{c}{Fe/O}  \\
 & (doy) & ave.&  max & ave. & max. & ave. & max. & ave. & max. \\
\tableline  

2002 & 173 & 416$\pm$19 & 464 & 4.4$\pm$1.1 & 6.9 & 80$\pm$10 & 92 &
0.15$\pm$0.03 & 0.19 \\

2000 & 210 & 442$\pm$34 & 474 & 15.1$\pm$7.5 & 35.9 &  227$\pm$245 &
1020 & 0.34$\pm$0.31 & 0.90 \\

2001 & 351 & 477$\pm$20 & 500 & 3.6$\pm$0.8 & 5.5 &  123$\pm$16 & 147 &
0.06 & 0.07 \\

2000 & 178 & 504$\pm$42 & 569 & 5.2$\pm$2.3 & 13.0 & 99$\pm$38 & 171 &
0.21$\pm$0.04 & 0.29 \\

1998 & 268 & 640$\pm$77 & 793 & 3.6$\pm$2.2 & 11.1 & 100$\pm$72 & 214 &
0.28$\pm$0.13 & 0.56 \\

2003 & 129 & 706$\pm$78 & 855 & 3.2$\pm$2.2 & 10.6 & 78$\pm$27 & 114 &
0.18$\pm$0.04 & 0.25 \\

2000 & 262 & 718$\pm$47 & 804 & 4.4$\pm$3.1 & 13.3 & 222$\pm$64 & 335 &
0.28$\pm$0.17 & 0.67 \\

2003 & 302 & 993$\pm$305 & 1700 & 3.1$\pm$1.9 & 9.2 &  168$\pm$129 & 442
& 0.67$\pm$0.42 & 2.33 \\

\tableline
\tablenotetext{a}{Ranges given are the standard deviation in the values
over the ICME event
                  and do not include uncertainties in the measurement.}
\tablenotetext{b}{Ratio of the number densities of the given elements}
\end{tabular}
\end{center}
\label{tab1}
\end{table}
\clearpage

\tabletypesize{\tiny}
\setlength{\tabcolsep}{0.3em}

\begin{table}[t]
\begin{center}
\caption{CME Models}
\begin{tabular}{@{}*{15}{c}}  
\tableline\tableline \\
Year & Day
& \multicolumn{6}{c}{Input Parameters}
& \multicolumn{4}{c}{Output Parameters} \\
 & & $v_{f}$ & $a$ & $v_{i}$ &
$\rho_{e}$ & $T_{e}$ & $\frac{QE}{KE}$ &
 $\rho_{10 R_{\sun}}$ &
\multicolumn{3}{c}{charge states}\\
 &  &
km s$^{-1}$ & km s$^{-2}$ &  km s$^{-1}$ &
$10^{7}$cm$^{-3}$ & $10^{6}$ K & &
$10^{3}$cm$^{-3}$ & O & Si & Fe \\
 & & \\
\tableline
2002 & 173 & 425 & 0.1 & 10 &  10 & 1 & 0.7 &  10 &
6 & 7,8,6 & 8,9  \\
2002 & 173 & 425 & 0.1 & 10 &  1 & 1 & 0 &  0.9 &
6 & 7,8,6 & 8,9,10 \\

2000& 210& 500& 0.1& 20&  10& 2& 5&  30& 7,6,8 & 12,11&
16,17,15\\
2000& 210& 500& 0.1& 20&  1& 1.3& 0&  2.9& 6 & 8,9,7&
11,8,12,10,9\\

2001& 351& 500-800& 0.1-0.25& 10&  5& 1.8& 6 &  7.0 & 6,7 & 12,11,10&
14,15,13,16\\
2001& 351& 500-800& 0.1-0.25& 10&  0.5& 1.2& 0&  0.8& 6 & 8,9,7&
11,10,9,8\\

2000 & 178 & 500-850 & 0.1-0.15 & 10 &  10 & 1 & 9 &  15 &
6,7 & 10,11,12 & 16,15,17 \\
2000 & 178 & 500-850 & 0.1-0.15 & 10 &  1 & 1 & 0 &  1.5 &
6 & 7,8,6 & 8,9,10 \\

1998 & 268 & 750 & 0.1 & 15 &  10 & 2.3 & 8 & 16 &
7,8 & 12,11 & 16,17 \\
1998 & 268 & 750 & 0.1 & 15 &  1 & 1 & 0 &  1.8 &
6 & 7,8 & 8,9,10 \\

2003 & 129 & 700 & 0.1 & 10 &  10 & 1.2 & 2.6 &
11 &
6 & 9,10,8 & 13,12,14,15 \\
2003 & 129 & 700 & 0.1 & 10 &  1 & 1.2 & 0 &
1.1 &
6 & 8,9,7 & 8,11,10,9 \\

2000 & 262 & 700-900 & 0.1-0.15 & 15 &  10 & 2.5 & 3-2.5 &  17
&
7,8,6 & 12,11 & 16,15,14 \\
2000 & 262 & 700-900 & 0.1-0.15 & 15 &  1 & 1.4 & 0 & 1.7
&
6 & 9,8,10 & 11,12,10,8 \\

2003 & 302 & 1300-2500 & 0.2-1 & 20 &  10 & 2.3 & 4-8  &
15&
7,8,6 & 12,11 & 16,17,15 \\
2003 & 302 & 1300-2500 & 0.2-1 & 20 &  1 & 1.2 & 0  &
1.4&
6 & 8,9,7 & 8,11,10,9 \\    

\tableline
\tablenotetext{a}{Charge states are listed in order of abundance composing a total charge fraction of at least 0.8.}
\end{tabular}
\end{center}
\label{tab2}
\end{table}


\begin{figure}
\includegraphics[width=3.0in]{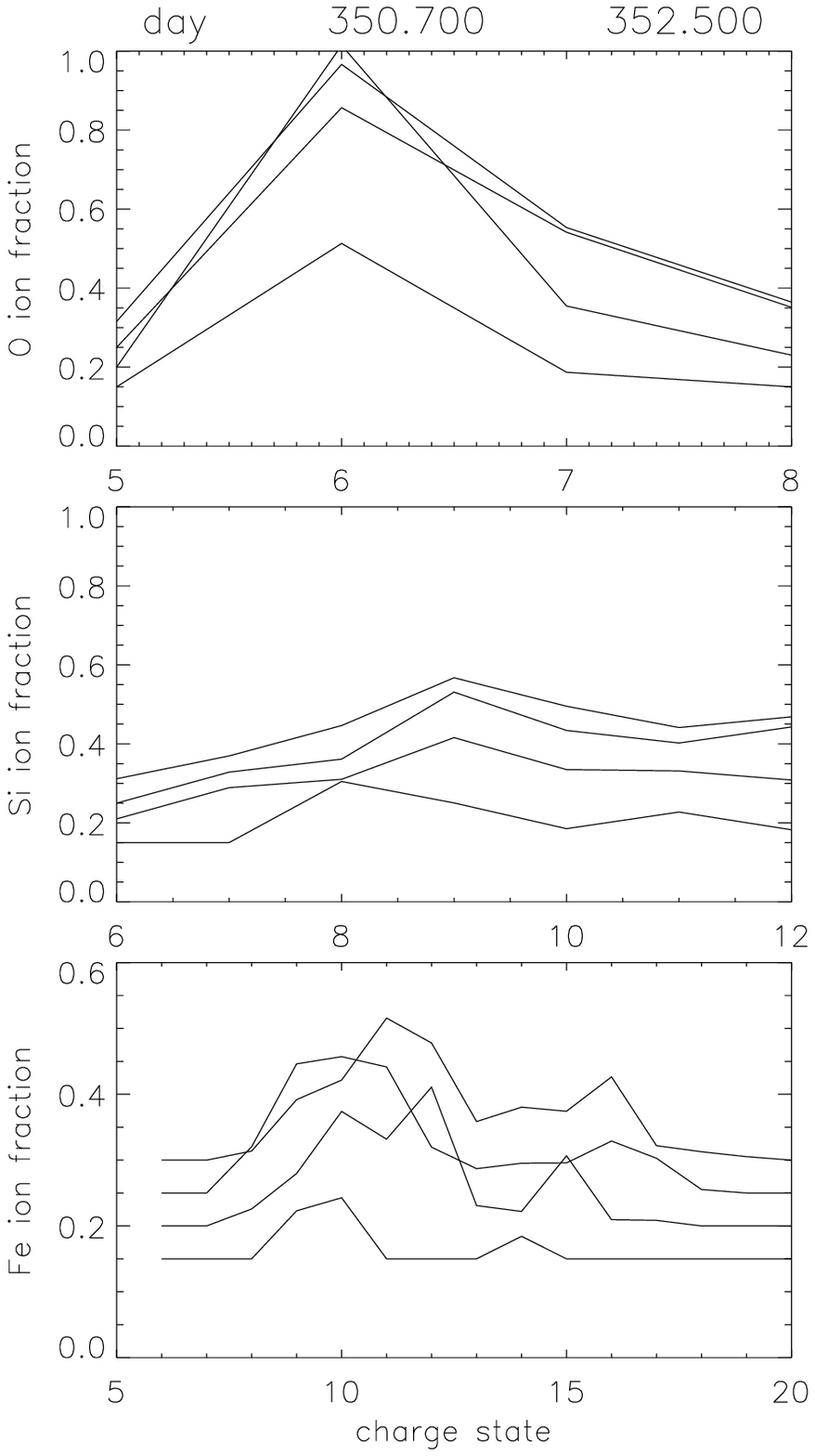}
\includegraphics[width=3.0in]{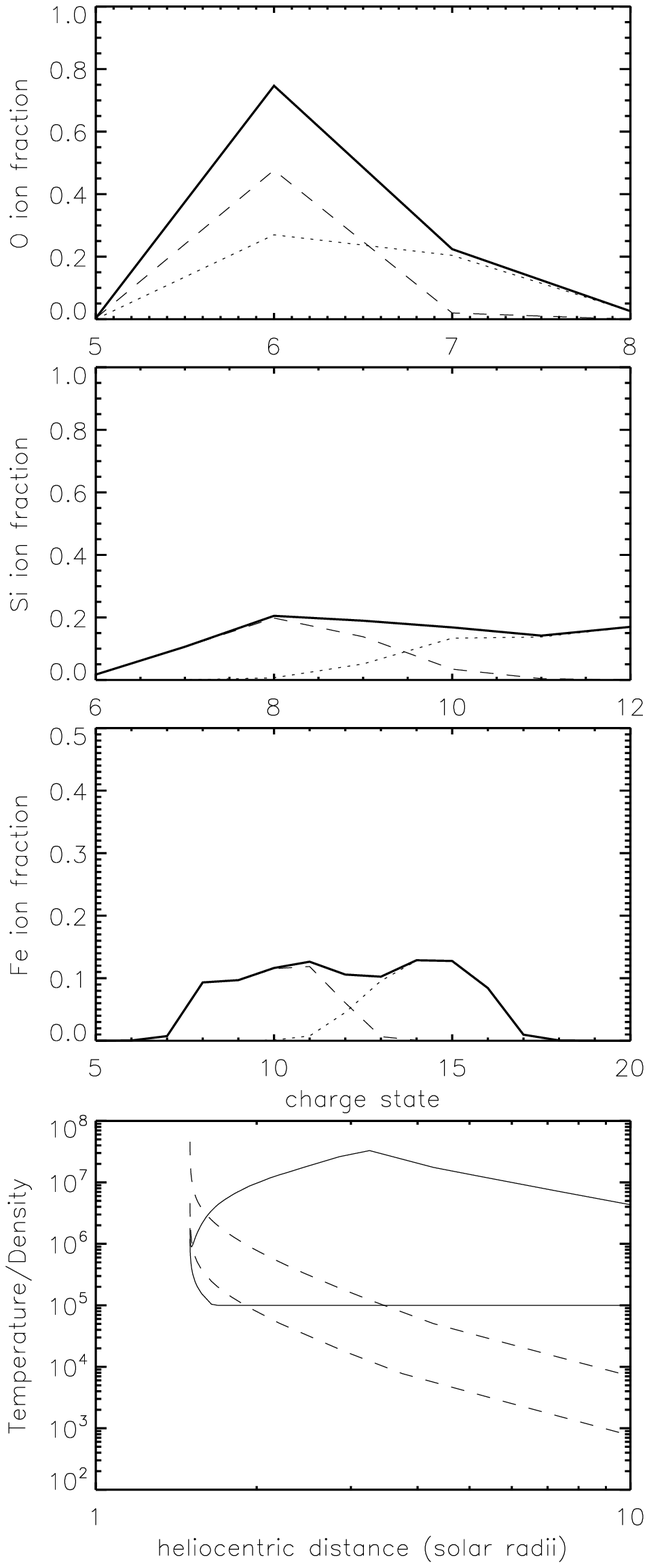}
\caption{Charge state distributions for O, Si, and Fe given in terms
of the ion fractions for the 2001 day 351 ICME. Plotted here are the
ten hour average distributions, offset by increments of 0.05. }
\caption{Model charge state distributions for the 2001 day 351 ICME
for parameters given in Table 2 with weighting of [0.5,0.5] for the
``core'' and ``cavity'' components respectively. The bottom panel
illustrates the temperature and density evolution close to the solar
surface.}
\end{figure}

\begin{figure}
\includegraphics[width=6.0in]{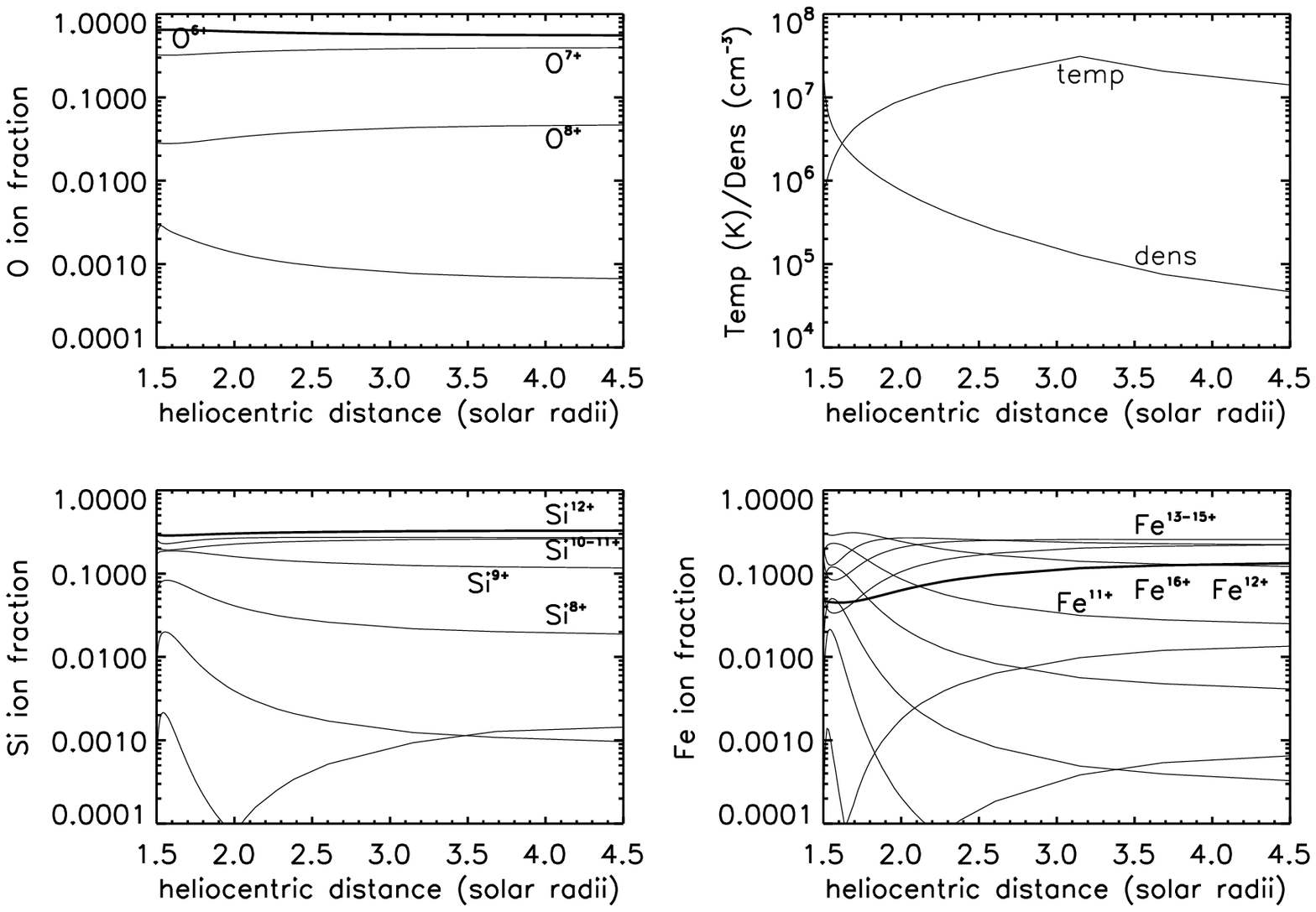}
\caption{Evolution of charge state distributions for O, Si, and Fe for the 2001 day 351 ``core'' for
parameters given in Table 2.
The upper right panel illustrates the temperature and density evolution close to the solar surface.}
\end{figure}


\begin{figure}
\includegraphics[width=3.0in]{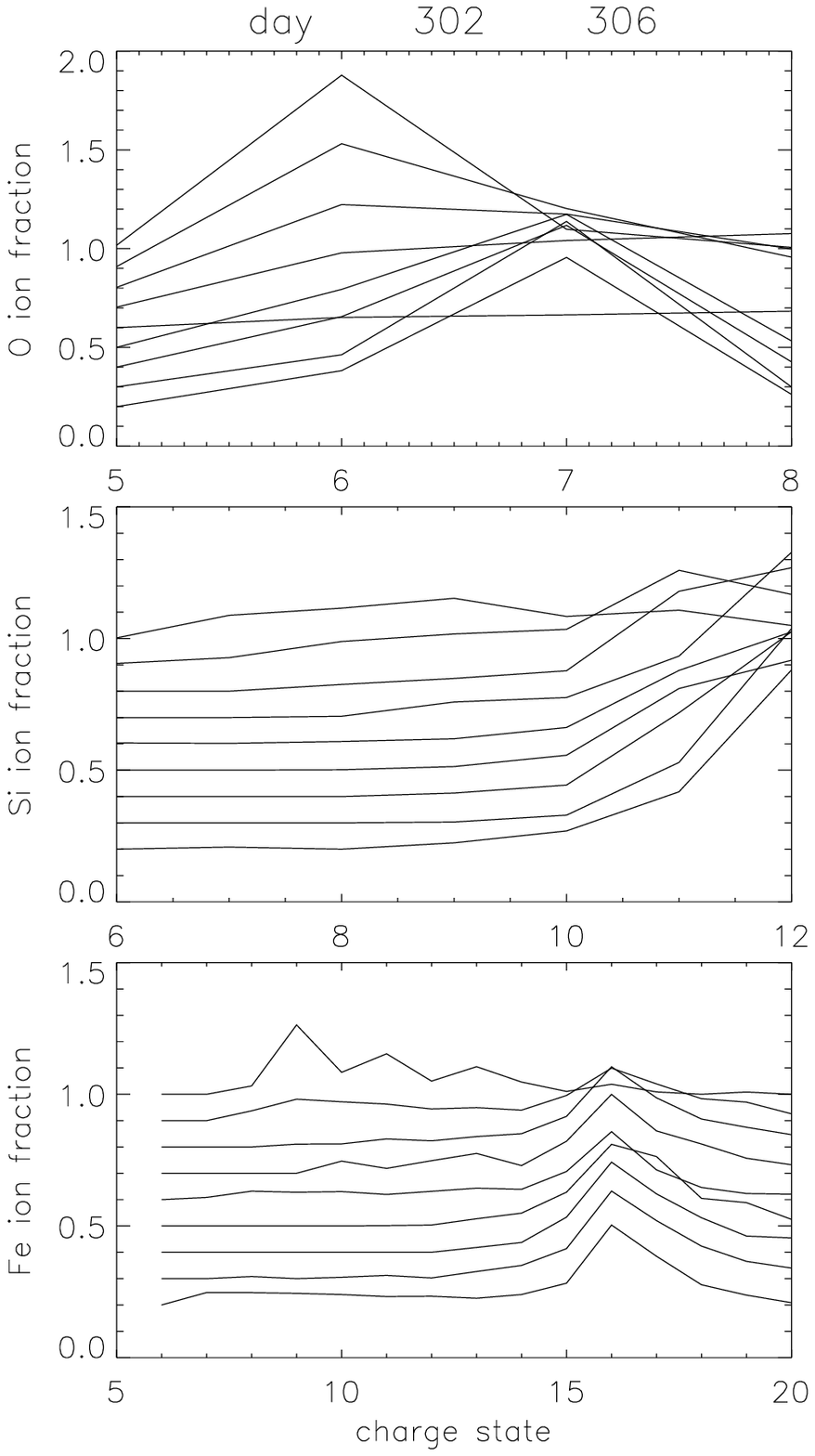}
\includegraphics[width=3.0in]{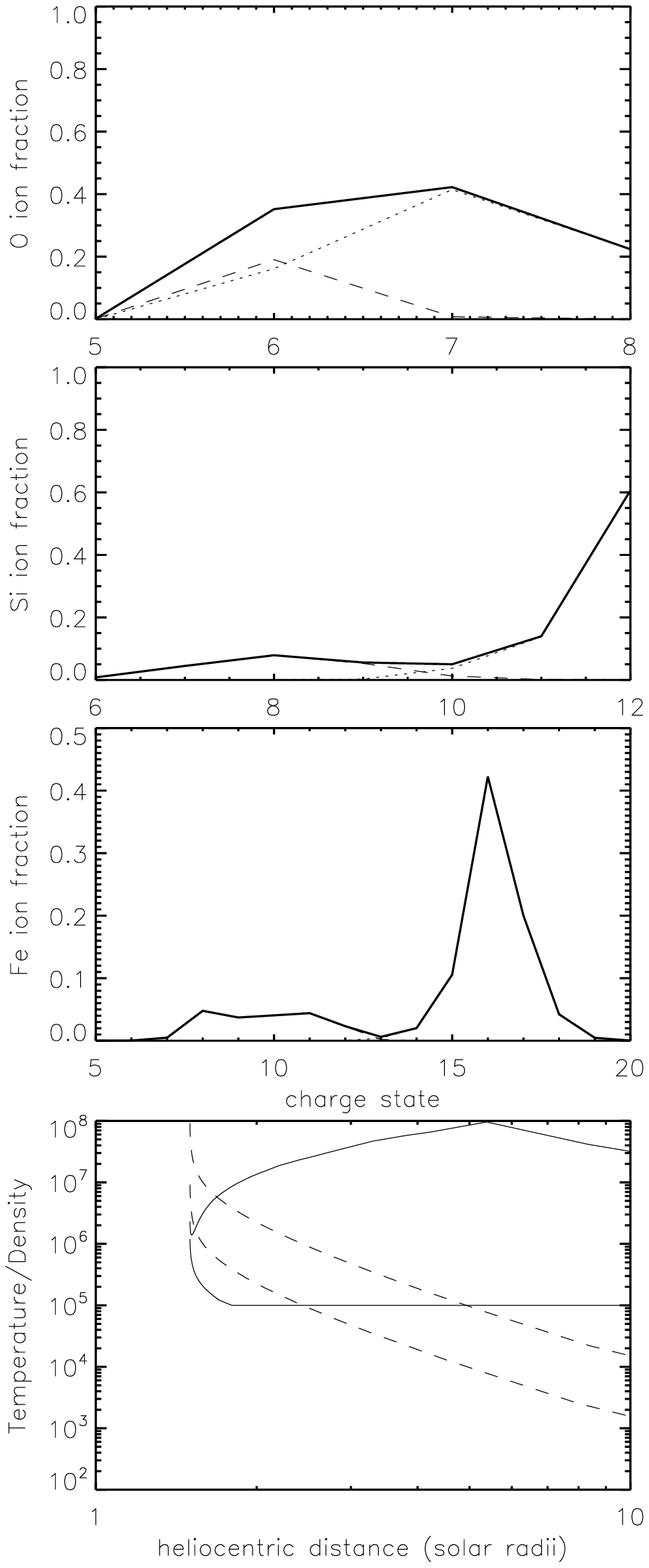}
\caption{Same as figure 3 for ICME 2003 day 302. Increment between ten hour averages is 0.1}
\caption{Same as figure 4 for ICME 2003 day 302, weighting [0.8,0.2] for the ``core'' and ``cavity'' contributions}
\end{figure}

\begin{figure}
\includegraphics[width=6.0in]{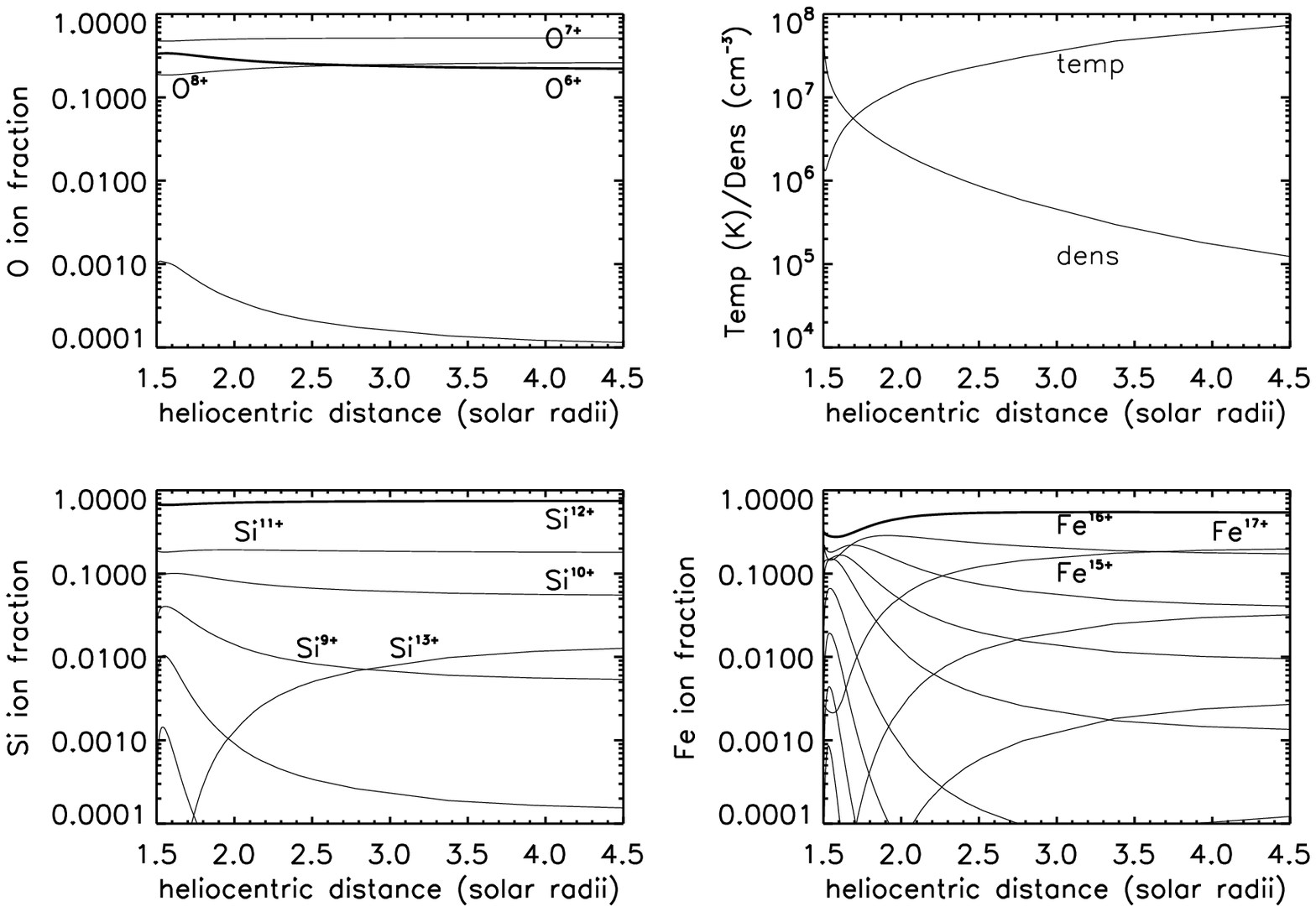}
\caption{Evolution of charge state distributions for O, Si, and Fe for the 2003 day 302 ``core'' for
parameters given in Table 2.
The upper right panel illustrates the temperature and density evolution close to the solar surface.}
\end{figure}



\begin{thebibliography}{}
\bibitem[Akmal et al.(2001)]{akmal01}Akmal, A., Raymond, J. C., Vourlidas, A., Thompson, B., Ciaravella,
A., Ko, Y.-K., Uzzo, M., \& Wu, R. 2001, \apj, 553, 922
\bibitem[Altun et al.(2004)]{altun04}Altun, Z., Yumak, A., Badnell, N. R.,
Colgan, J., \& Pindzola, M. S. 2004, \aap, 420, 775
\bibitem[Antiochos(1998)]{antiochos98}Antiochos, S. K. 1998, \apj, 502, L181
 \bibitem[Antiochos et al.(1999)]{antiochos99}Antiochos, S. K., DeVore, C. R., \& Klimchuk, J. A.
1999, \apj, 510, 485
\bibitem[Aulanier et al.(2002)]{aulanier02}Aulanier, G., DeVore, C. R., \& Antiochos, S. K. 2002,
\apj, 567, L97
\bibitem[Badnell(2006)]{badnell06}Badnell, N. R. 2006, J. Phys. B., 39, 4825
\bibitem[Ciaravella et al.(2001)]{ciaravella01}Ciaravella, A., Raymond, J. C., Reale, F., Strachan, L.,
\& Peres, G. 2001, \apj, 557, 351
\bibitem[Colgan et al.(2003)]{colgan03}Colgan, J., Pindzola, M. S., Whiteford,
A. D., \& Badnell, N. R. 2003, \aap, 412, 597
\bibitem[Colgan, Pindzola, \& Badnell(2004)]{colgan04}Colgan, J., Pindzola,
M. S., \& Badnell, N. R. 2004, \aap, 417, 1183
\bibitem[Dasgupta \& Whitney(2004)]{dasgupta04}Dasgutpa, A., \& Whitney, K. G.
2004, PRA, 69, 022702
\bibitem[Dere et al.(1997)]{dere97}Dere, K. P., et al. 1997, Solar Physics, 175, 601
\bibitem[DeVore \& Antiochos(2000)]{devore00}DeVore, C. R., \& Antiochos, S. K. 2000, \apj, 539, 954
\bibitem[Feldman et al.(2005)]{feldman05}Feldman, U., Landi, E., \& Laming, J. M. 2005, \apj, 619, 1142
\bibitem[Forbes(1991)]{forbes91}Forbes, T. G. 1991, Geophys. Astrophys. Fluid Dyn., 62, 15
\bibitem[Galvin \& Gloeckler(1997)]{galvin97}Galvin, A., \& Gloeckler, G. 1997,
in Correlated Phenomena at the Sun, in the Heliosphere and in
Geospace, ed. A. Wilson, 323
\bibitem[Gloeckler et al.(1998)]{gloeckler98}Gloeckler, G. et al. 1998, Space Science Reviews, 86, 497
\bibitem[Gosling et al.(1974)]{gosling74}Gosling, J. T., Hildner, E., MacQueen, R. M., Munro, R. H.,
 Poland, A. I., \& Ross, C. L. 1974, J. Geophys. Res., 79, 4581
\bibitem[Gu(2003)]{gu03}Gu, M. F. 2003, \apj, 590, 1131
\bibitem[Gu(2004)]{gu04}Gu, M. F. 2004, \apjs, 153, 389
\bibitem[Henke et al.(2001)]{henke01}Henke, T., Woch, J., Schwenn, R., Mall, U., Gloeckler, G.,
von Steiger, R., Forsyth, R. J., \& Balogh, A.. 2001, JGR, 106, 10597
\bibitem[Henke et al.(1998)]{henke98}Henke, T., Woch, J., Mall, U., Livi, S., Wilken, B., Schwenn, R.,
Gloeckler, G., von Steiger, R., Forsyth, R. J., \& Balogh, A. 1998, Geophys. Res. Lett., 25, 3465
\bibitem[Howard et al.(1985)]{howard85}Howard, R. A., Sheeley, N. R., Michels, D. J., \& Koomen, M. J.
1985, J. Geophys. Res., 90, 8173
\bibitem[Kasper et al.(2007)]{kasper07}Kasper, J. C., Stevens, M.
L., Lazarus, A. J., Steinberg, J. T., \& Ogilvie, K. W. 2007, \apj,
660, 901
\bibitem[Klassen et al.(2005)]{klassen05}Klassen, A., Krucker, S., Kunow, H., M\"uller-Mellin, R.,
Wimmer-Schweingruber, R., Mann, G., \& Posner, A. 2005, JGR, 110, A09S04
\bibitem[Kumar \& Rust(1996)]{kumar96}Kumar, A., \& Rust, D. M. 1996, JGR, 101, 15667
\bibitem[Laming(2004)]{laming04}Laming, J. M. 2004, \apj, 604, 874
\bibitem[Laming \& Feldman(2003)]{lamingf03}Laming, J. M., \& Feldman, U. 2003, \apj, 591, 1257
\bibitem[Laming \& Hwang(2003)]{laming03}Laming, J. M., \& Hwang, U. 2003, \apj, 597, 347
\bibitem[Laming \& Grun(2002)]{laming02}Laming, J. M., \& Grun, J. 2002, Phys. Rev.
Lett., 89, 125002
\bibitem[Laming \& Grun(2003)]{laming03g}Laming, J. M., \& Grun, J. 2003, Phys. Plasmas, 10, 1614
\bibitem[Lee et al.(2007)]{lee07}Lee, J.-Y., Raymond, J. C., Ko, Y.-K., \& Kim, K.-S. 2007, in preparation
\bibitem[Lepri \& Zurbuchen(2004)]{lepri04}Lepri, S. T., \&
Zurbuchen, T. H. 2004, JGR, 109, 1112
\bibitem[Lepri et al.(2001)]{lepri01}Lepri, S. T., Zurbuchen, T. H., Fisk, L. A., Richardson, I. G.,
Cane, H. V., \& Gloeckler, G. 2001, JGR, 106, 29231
\bibitem[Lin \& Forbes(2000)]{lin00}Lin, J., \& Forbes, T. G. 2000, JGR, 105, 2375
\bibitem[Lynch et al.(2004)]{lynch04}Lynch, B. J., Antiochos, S. K., MacNeice, P. J.,
Zurbuchen, T. H., \& Fisk, L. A. 2004, \apj, 617, 589
\bibitem[Lynch(2006)]{lynch06}Lynch, B. J. 2006, PhD Thesis, University of Michigan
\bibitem[Mann et al.(2003)]{mann03}Mann, G., Klassen, A., Aurass, H., \& Classen, H.-T. 2003, \aap,
400, 329
\bibitem[Mazzotta et al.(1998)]{mazzotta98} Mazzotta, P., Mazzitelli, G.,
Colafrancesco, S., \& Vittorio, N. 1998, \aaps, 133, 403
\bibitem[Mewaldt et al.(2005)]{mewaldt05}Mewaldt, R. A., Cohen, C. S., Labrador, A. W., Leske, R. A.,
Mason, G. M., Desai, M. I., Looper, M. D., Mazur, J. E., Selesnick, R. S., \& Haggerty, D. K. 2005,
JGR, 110, A09S18
\bibitem[Mitnik \& Badnell(2004)]{mitnik04}Mitnik, D. M., \& Badnell, N. R.
2004, \aap, 425, 1153
\bibitem[Reinard(2005)]{reinard05}Reinard, A. 2005, \apj, 620, 501
\bibitem[Riley et al.(2007)]{riley07}Riley, P., Lionello, R., Miki\'c, Z., Linker, J., Clark, E.,
Lin, J., \& Ko, Y.-K. 2007, \apj, 655, 591
\bibitem[Riley \& Crooker(2004)]{riley04}Riley, P., \& Crooker, N. U. 2004, \apj, 600, 1035
\bibitem[Riley et al.(2002)]{riley02}Riley, P., Linker, J. A, Miki\'c, Z., Odstrcil, D., Pizzo, V. J.,
\& Webb, D. F. 2002, \apj, 578, 972
\bibitem[Schmidt \& Cargill(2001)]{schmidt01}Schmidt, J. M., \& Cargill, P. J. 2001, JGR, 106, 8283
\bibitem[Shiota et al.(2005)]{shiota05}Shiota, D., Isobe, H., Chen, P. F., Yamamoto, T. T., Sakajiri, T.,
\& Shibata, K. 2005, \apj, 634, 663
\bibitem[Thernisien et al(2006)]{thernisien06}Thernisien, A. F. R., Howard, R. A., \& Vourlidas,
A. 2006, \apj, 652, 763
\bibitem[Ugarte-Urra et al.(2007)]{ugarte07}Ugarte-Urra, I., Warren, H. P., \& Winebarger, A. R. 2007,
\apj, in press
\bibitem[Vourlidas et al.(2003)]{vourlidas03}Vourlidas, A., Wu, S. T., Wang, A. H.,
Subramanian, P., \& Howard, R. A. 2003, \apj, 598, 1392
\bibitem[Zatsarinny et al.(2004a)]{zatsarinny04a}Zatsarinny, O., Gorczyca, T.
W., Korista, K. T., Badnell, N. R., \& Savin. D. W. 2004, \aap, 417, 1173
\bibitem[Zatsarinny et al.(2003)]{zatsarinny03}Zatsarinny, O., Gorczyca, T. W.,
Korista, K. T., Badnell, N. R., \& Savin. D. W. 2003, \aap, 412, 587
\bibitem[Zatsarinny et al.(2004b)]{zatsarinny04b}Zatsarinny, O., Gorczyca, T.
W., Korista, K. T., Badnell, N. R., \& Savin. D. W. 2004, \aap, 426, 699
\bibitem[Zhang et al.(2001)]{zhang01}Zhang, J., Dere, K. P., Howard, R. A., Kundu, M. R., \&
White, S. M. 2001, \apj, 559, 452
\bibitem[Zhang et al.(2004)]{zhang04}Zhang, J., Dere, K. P., Howard, R. A., \& Vourlidas, A. 2004,
\apj, 604, 420
\bibitem[Zhang \& Dere(2006)]{zhang06}Zhang, J., \& Dere, K. P. 2006, \apj, 649, 1100
\bibitem[Zurbuchen et al.(2004)]{zurbuchen04} Zurbuchen, T.~H.,
Gloeckler, G., Ipavich, F., Raines, J., Smith, C.~W., \& Fisk, L.~A.\ 2004,
\grl, 31, 11805
\bibitem[Zurbuchen \& Richardson(2006)]{zurbuchen06}Zurbuchen, T.
H., \& Richardson, I. G. 2006, Space Science Reviews, 123, 31

\end{thebibliography}
\end{document}